\documentclass[review,3p]{elsarticle}
\usepackage{lineno,hyperref}
\pdfoutput=1
\usepackage{color}

\journal{Astroparticle physics Journal}

\begin{document}

\begin{frontmatter}

\title{First results from the NEWS-G direct dark matter search experiment at the LSM}

\author[a]{Q.~Arnaud\corref{mycorrespondingauthor}}
\cortext[mycorrespondingauthor]{Corresponding author: q.arnaud@queensu.ca}
\author[b]{D.~Asner}
\author[c]{J.-P.~Bard}
\author[a,c]{A.~Brossard}
\author[a]{B.~Cai} 
\author[a]{M.~Chapellier}
\author[a]{M.~Clark}
\author[d]{E.~C.~Corcoran}
\author[e]{T.~Dandl}
\author[f]{A.~Dastgheibi-Fard}
\author[a]{K.~Dering}
\author[a]{P.~Di Stefano}
\author[a]{D. Durnford}
\author[a]{G.~Gerbier}
\author[c]{I.~Giomataris}
\author[g]{P.~Gorel}
\author[c]{M.~Gros}
\author[h]{O.~Guillaudin}
\author[b]{E.~W.~Hoppe}
\author[a]{A.~Kamaha} 
\author[c]{I.~Katsioulas}
\author[d]{D.~G.~Kelly}
\author[a]{R.~D. Martin}
\author[a]{J.~McDonald}
\author[h]{J.-F.~Muraz}
\author[c]{J.-P.~Mols}
\author[c]{X.-F.~Navick}
\author[c]{T.~Papaevangelou}
\author[f]{F.~Piquemal}
\author[a]{S.~Roth\fnref{myfootnotesabine}}
\fntext[myfootnotesabine]{Now at Technical University of Munich}
\author[h]{D.~Santos}
\author[i]{I.~Savvidis}
\author[e]{A.~Ulrich}
\author[a]{F.~Vazquez de Sola Fernandez} 
\author[f]{M.~Zampaolo}



\address[a]{Department of Physics, Engineering Physics \& Astronomy, Queen's University, Kingston, Ontario K7L 3N6, Canada}
\address[b]{Pacific Northwest National Laboratory, Richland, Washington 99354, USA}
\address[c]{IRFU, CEA, Universit\'{e} Paris-Saclay, F-91191 Gif-sur-Yvette, France}
\address[d]{Chemistry \& Chemical Engineering Department, Royal Military
College of Canada, Kingston, Ontario K7K 7B4, Canada}
\address[e]{Physik Department E12, Technische Universit\"{a}t M\"{u}nchen, James-Franck-Str. 1, 85748 Garching, Germany}
\address[f]{LSM, CNRS/IN2P3, Universit\'{e} Grenoble-Alpes, Modane, France }
\address[g]{SNOLAB, Lively, Ontario, P3Y 1N2, Canada}
\address[h]{LPSC, Universit\'{e} Grenoble-Alpes, CNRS/IN2P3, Grenoble, France}
\address[i]{Aristotle University of Thessaloniki, Thessaloniki, Greece}

\begin{abstract}
New Experiments With Spheres-Gas (NEWS-G) is a direct dark matter detection experiment using Spherical Proportional Counters (SPCs) with light noble gases to search for low-mass Weakly Interacting Massive Particles (WIMPs). We report the results from the first physics run taken at the Laboratoire Souterrain de Modane (LSM) with SEDINE, a 60 cm diameter prototype SPC operated with a mixture of Ne + $\mathrm{CH}_{4}$ (0.7~\%) at 3.1 bars for a total exposure of $9.7\;\mathrm{kg\cdot days}$. New constraints are set on the spin-independent WIMP-nucleon scattering cross-section in the sub-$\mathrm{GeV/c^2}$ mass region. We exclude cross-sections above $4.4 \times \mathrm{10^{-37}\;cm^2}$ at 90~\% confidence level (C.L.) for a 0.5 $\mathrm{GeV/c^2}$ WIMP. The competitive results obtained with SEDINE are promising for the next phase of the NEWS-G experiment: a 140 cm diameter SPC to be installed at SNOLAB by summer 2018. 
\end{abstract}


\end{frontmatter}

\section{Introduction}
It is now well established, from a wide variety of astrophysical observations~\cite{Bertone2010,bullet} and precise measurements of the Cosmological Microwave Background (CMB)~\cite{Planck2015}, that non-baryonic cold dark matter is an essential ingredient to our understanding of the Universe. Although the nature of dark matter remains unknown, many theories beyond the Standard Model of particle physics predict massive and neutral particles, thermally produced in the early Universe, that may account for the observed relic density: a generic class of well motivated dark matter candidates known as Weakly Interacting Massive Particles (WIMPs)~\cite{DMCANDIDATES}. Direct detection experiments aim to detect incoming WIMPs from the Milky Way halo via their coherent elastic scattering off of target nuclei. Expected event rates are {orders of magnitude lower than natural radioactivity, requiring large exposures with detectors made of radio-pure materials, shielded and operated deep underground to protect against ambient radioactivity and cosmic rays, respectively. Additional experimental constraints come from the exponential shape of the expected nuclear recoil energy spectrum, whose slope gets steeper and whose endpoint quickly drops as the WIMP mass, $m_\chi$, decreases. Low energy detection thresholds ($<$10 keV) are therefore critical, especially to be sensitive to low-mass WIMPs. Originally, WIMP masses in the range of $\mathrm{10\;GeV/c^2}$ to $\mathrm{1\;TeV/c^2}$ were favoured by most supersymmetric (SUSY) models~\cite{JUNGMAN}. Large dual phase Time Projection Chambers (TPCs)~\cite{List-Limits-lux2016.txt,PANDAX} are particularly well suited for this mass range, and currently dominate other detection techniques in the spin-independent sector. However, lack of evidence for SUSY at LHC~\cite{PaperLHC}, and failure from direct detection experiments to find any candidate at masses above $\mathrm{10\;GeV/c^2}$, suggest that the search should be extended, in particular to lower mass candidates. Indeed, several new theoretical approaches such as dark sector~\cite{darksector,goodnews}, asymmetric dark matter \cite{ReviewAsymDarkMatter2013,AsymDarkMatterSignaturesAndConstraints2013} and generalized effective theory~\cite{EFFECTIVEFIELD} pave the way for candidates with lower mass and/or more complex couplings than the traditional spin (in)dependent ones. Furthermore, upper-limits set on the spin-independent WIMP-nucleon scattering cross section $\sigma_{SI}$ in the $\mathrm{\;GeV/c^2}$ mass range are currently orders of magnitude weaker than those set at higher mass~\cite{List-Limits-cresst2015.txt,List-Limits-supercdms2014.txt} because threshold requirements overwhelm all other experimental constraints. Thus, a wide region of the parameter space ($\sigma_{SI}$,$m_{\chi}$) can be probed with low energy detection thresholds without relying on tonne-scale absorbers. 

In this context, the New Experiments With Spheres-Gas (NEWS-G)~\cite{newsgwebsite} project benefits from key features arising from the novel technology of Spherical Proportional Counters (SPCs)~\cite{ioanis,Ajout1,Ajout2,Ajout3,Ajout4}} (see Sec~\ref{sec:ExperimentalSetUp}): first, the high amplification gain combined with the low intrinsic capacitance of the central sensor ($\sim0.35\;\mathrm{pF}$) results in a signal to noise ratio that provides sensitivity to single electrons from primary ionization and allows for extremely low detection thresholds of 10 to 40 $\mathrm{eV_{ee}}$ (electron equivalent recoil energy). Second, the use of light target gases such as H, He and Ne allows for the optimization of momentum transfers for low-mass particles in the $\mathrm{GeV/c^2}$ mass range, significantly increasing the sensitivity to sub-$\mathrm{GeV/c^2}$ WIMPs. Finally, the possibility to operate with targets of diverse atomic numbers, $A$, allows for the discrimination of a potential WIMP signal from a significant unidentified background based on the $A^2$ dependence of the coherent WIMP-nucleon scattering cross-section.   

In this paper, we report on the analysis of a 42.7 day-long physics-run taken between April and May 2015 at the Laboratoire Souterrain de Modane with a prototype SPC called SEDINE. This detector was built to assess the relevance of SPCs for dark matter searches. The contamination in radioactive elements of the materials used for SEDINE was not controlled precisely enough to interpret a potential excess of events above expected backgrounds as a WIMP signal. Therefore, it was decided prior to analysing the data that the results would be reported in terms of an upper limit on the WIMP-nucleon cross-section at 90~\% confidence level (C.L.).
 
This paper is organized as follows: we first describe in section \ref{sec:ExperimentalSetUp} the experimental set-up and provide the reader with a comprehensive description of both the physics of SPCs and our treatment of pulses recorded with the detector. In section \ref{sec:simulations}, we detail the full simulation of the response of the detector, which is used to compute our sensitivity, and show its good agreement with calibration data. In section \ref{sec:dataanalysis}, we present the multivariate analysis of the physics-run which is based on a Boosted Decision Tree (BDT)~\cite{BDT} algorithm to optimize our signal/background discrimination. The competitive results for low-mass WIMPs reported in section \ref{sec:results} are promising for the next phase of the NEWS-G experiment which is presented in section \ref{sec:conclusion} with an overview of the major improvements.

\section{Detector principle, experimental set-up and operating conditions}
\label{sec:ExperimentalSetUp}
SEDINE is installed at the Laboratoire Souterrain de Modane (LSM) under a rock thickness of 4800~mwe. The detector is additionally protected from external radiation by a cubic shielding made of, from the inside to the outside, 8~cm of copper, 15~cm of lead and 30~cm of polyethylene. 
SEDINE consists of a 60 cm diameter sphere made of ultra pure (NOSV) copper~\cite{NOSV} and a 6.3~mm diameter spherical sensor made of silicon located at the center of the vessel. This small sensor was biased to 2520~V via a 380~$\mathrm{\mu m}$ diameter insulated HV wire routed through a grounded copper rod. The sphere was filled with a mixture of neon (99.3~\% in pressure) and CH${_4}$ (0.7~\%) at a total pressure of 3.1 bar and operated in sealed mode for 42.7 days without interruption.

\begin{figure}
\begin{minipage}{0.5\linewidth}
\centerline{\includegraphics[width=0.8\linewidth]{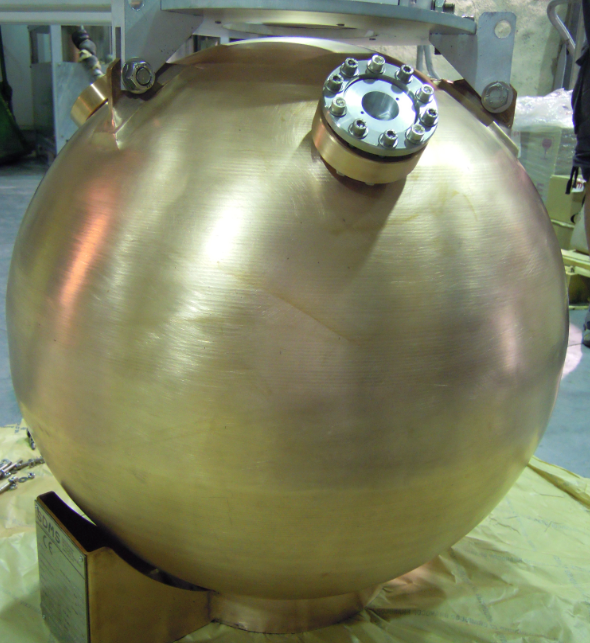}}
\end{minipage}
\begin{minipage}{0.57\linewidth}
\centerline{\includegraphics[width=0.95\linewidth]{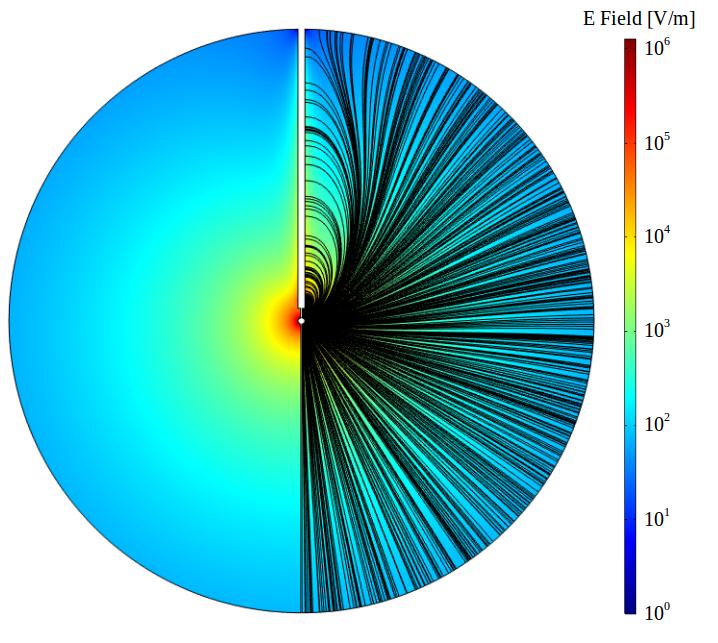}}
\end{minipage}
\caption{ Left: Picture of SEDINE: a 60 cm diameter prototype made of low activity (NOSV) copper. Right: Cross section of the detector in axial symmetry. The rod and the sensor are visible in white at scale. Field lines are shown on the right part of the picture only to allow the electric-field-magnitude values indicated in log scale by the colour code to be properly seen.}
\label{fig:detector}
\end{figure}

We show in Fig.~\ref{fig:detector} (left) a picture of SEDINE, together with the configuration of the electric field in Fig.~\ref{fig:detector} (right) for a cross section of the detector. The electric field magnitude depends on $1/\mathrm{r^2}$ in most of the volume, where it is unaffected by the presence of the grounded rod. Following an ionizing energy deposition, primary electrons (PEs) from ionization drift towards the sensor (anode) within hundreds of $\mathrm{\mu s} $ in the low field region where they undergo diffusion. As a result, the arrival time of the PEs on the sensor is subject to a Gaussian dispersion whose standard deviation, later referred to as ``diffusion time'', approximatively increases with the radial distance of the energy deposit as $\sigma(r)=(\frac{r}{r_{sphere}})^3\times 20\;\mathrm{\mu s} $. 
When the PEs reach the high field region, within less than $1\;\mathrm{mm}$ of the sensor, they gain enough kinetic energy between collisions with gas molecules to ionize the gas and create thousands of secondary electron/ion pairs. 
As the secondary ions drift away from the sensor, they induce a current which is integrated by a charge sensitive preamplifier and then digitized at 2.08 MHz. The detector response $D(t)$ to a single PE event is therefore a convolution of the current induced by the ions on the sensor and of the delta response of the preamplifier. Because each of the $\mathrm{N_{PE}}$ PEs arrives at the avalanche region at a particular time $t_i$ and leads to a different number $N_i$ of secondary ionizations, the acquired signal can be expressed as follows: $S(t)=\sum  _{i=1}^{\mathrm{N_{PE}}}N_i\times D(t-t_i)$.
Despite ions drifting towards the sphere at ground (cathode) within seconds, more than 50~\% of the signal is induced within the first $30 \; \mathrm{\mu s}$ because their speed decreases as $1/r^2$. Still, the fast decay time constant ($\tau=50 \; \mathrm{\mu s}$) of our resistive feedback charge sensitive preamplifier (CANBERRA Model 2006) is responsible for a loss in the pulse-height. The latter effect, called  ballistic deficit, increases with the diffusion time of the PEs. As a consequence, the raw pulse amplitude not only depends on the energy but also on the initial location of the event. To correct for this effect, one can separately deconvolve the raw pulse by the preamplifier response and by the ion induced current to determine the amplitude from the integral of the deconvolved pulse.
However, this double deconvolution procedure can greatly amplify high frequency noise and degrade the energy resolution. To avoid this effect and still correct efficiently for ballistic deficit, in the present analysis raw-pulses were only deconvolved once using a single effective exponential decay, and run through a low-pass filter. The associated time constant of this ad hoc detector response was chosen to better approximate both the preamplifier and the ion induced current responses at once. 

\begin{figure}[t!]
 \centering
\includegraphics[width=1\linewidth]{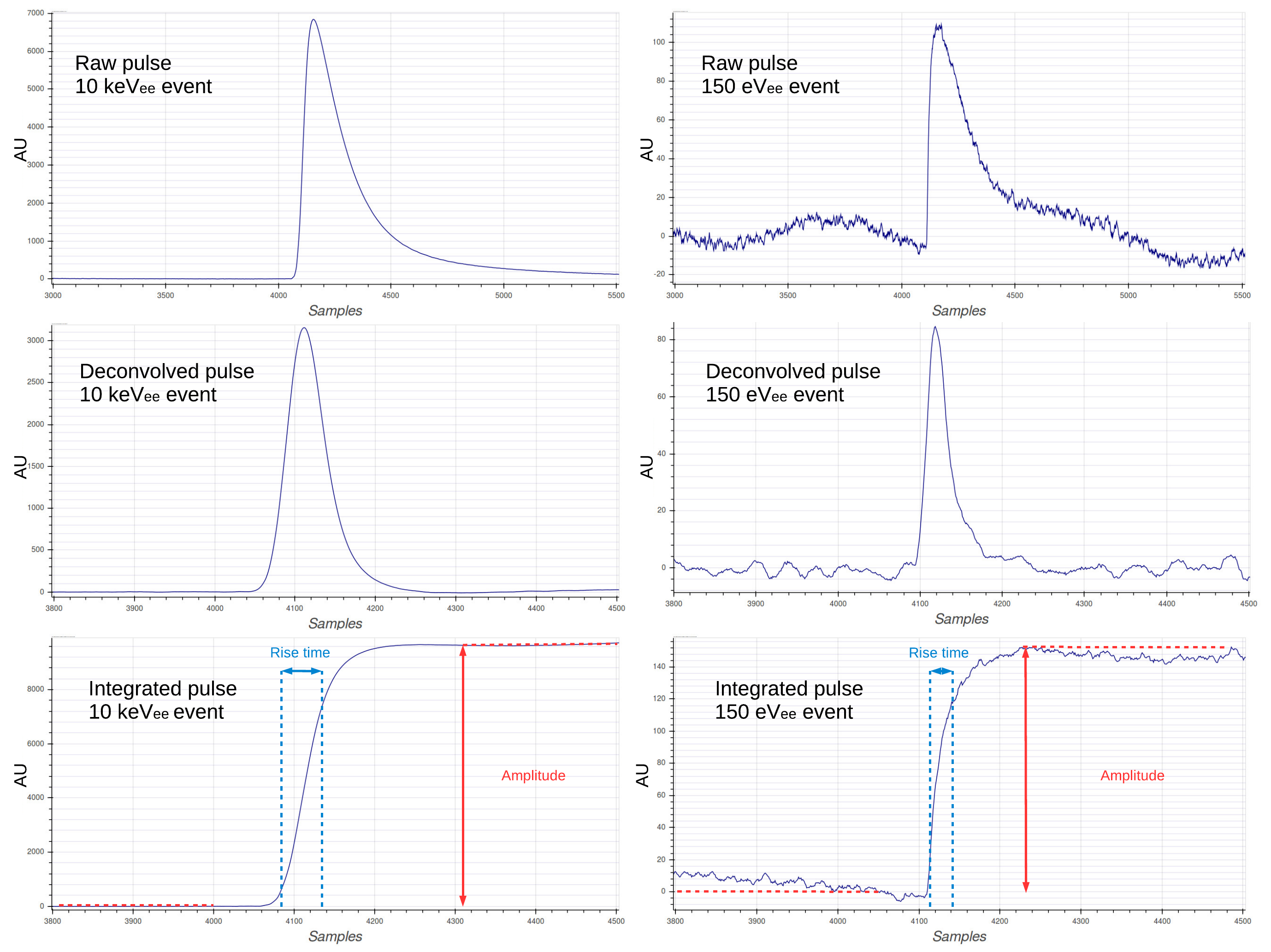}
\caption{Pulse treatment applied to a 10 $\mathrm{keV_{ee}}$ event (left panels) and to a 150 $\mathrm{eV_{ee}}$ event (right panels) recorded during the physics-run. Top pannels: raw pulses in a 1250 $\mathrm{\mu s}$ time window (1 sample = 480 ns). Middle: zoom in a 350 $\mathrm{\mu s}$ time window of the deconvolved pulse. Bottom: Integrated charge pulse from which the amplitude and rise time are extracted. It is obtained by successively applying a low-pass filter to the deconvolved pulse and performing a cumulative integration.}
\label{fig:pulse}
\end{figure}

We show in Fig.~\ref{fig:pulse} the pulse treatment discussed above, applied to a 10 $\mathrm{keV_{ee}}$ event (left panels) and to a 150 $\mathrm{eV_{ee}}$ event (right panels) recorded during the physics-run. Raw pulses are shown on top panels while the deconvolved pulses and their cumulative integration are shown on the middle and bottom panels, respectively. The two main analysis parameters that are extracted from the treated pulse are the amplitude, now proportional to the deposited energy only, and the rise time, defined as the time it takes to go from 10~\% to 75~\% of the amplitude of the integrated charge pulse. The rise time increases with the diffusion time and consequently, with the radial location, $r$, of the event. For events with a large number of PEs (c.f.~Fig.~\ref{fig:pulse} (left)), the deconvolved pulse matches the Gaussian distribution of the arrival times of the PEs. In that case, the rise time increases with the diffusion time $\sigma(r)$ as $\mathrm{RT}(r)=1.96\times \sigma(r)$ since it corresponds to the integral from the 10~\% to the 75~\% quantile of a Gaussian distribution with standard deviation  $\sigma(r)$.

The rise time is a key parameter of the analysis as it allows us to discriminate between surface events, mostly originating from radioactive contamination of the inner surface of the sphere, and WIMP events, expected to be uniformly distributed in the vessel. Surface events are our dominating background at low energy. These are expected from radon daughter decays, mostly $\beta-$rays from $^{210}$Pb decay and Auger electrons from the de-excitation of $^{210}\mathrm{Bi^{*}}$, both of which are of a few tens of keV, therefore being stopped within less than 1 cm in 3.1 bars of Ne. Another anticipated background consists of Compton/photoelectric interactions expected to be homogeneously distributed in the volume. These mainly originate from high energy $\gamma-$rays from $^{208}$Tl and $^{40}$K present in the rock, and from the decay chains of $^{238}$U and $^{232}$Th contained in the copper shell and the shielding itself.  
\begin{figure}[b!]
\centering
\includegraphics[width=0.8\linewidth]{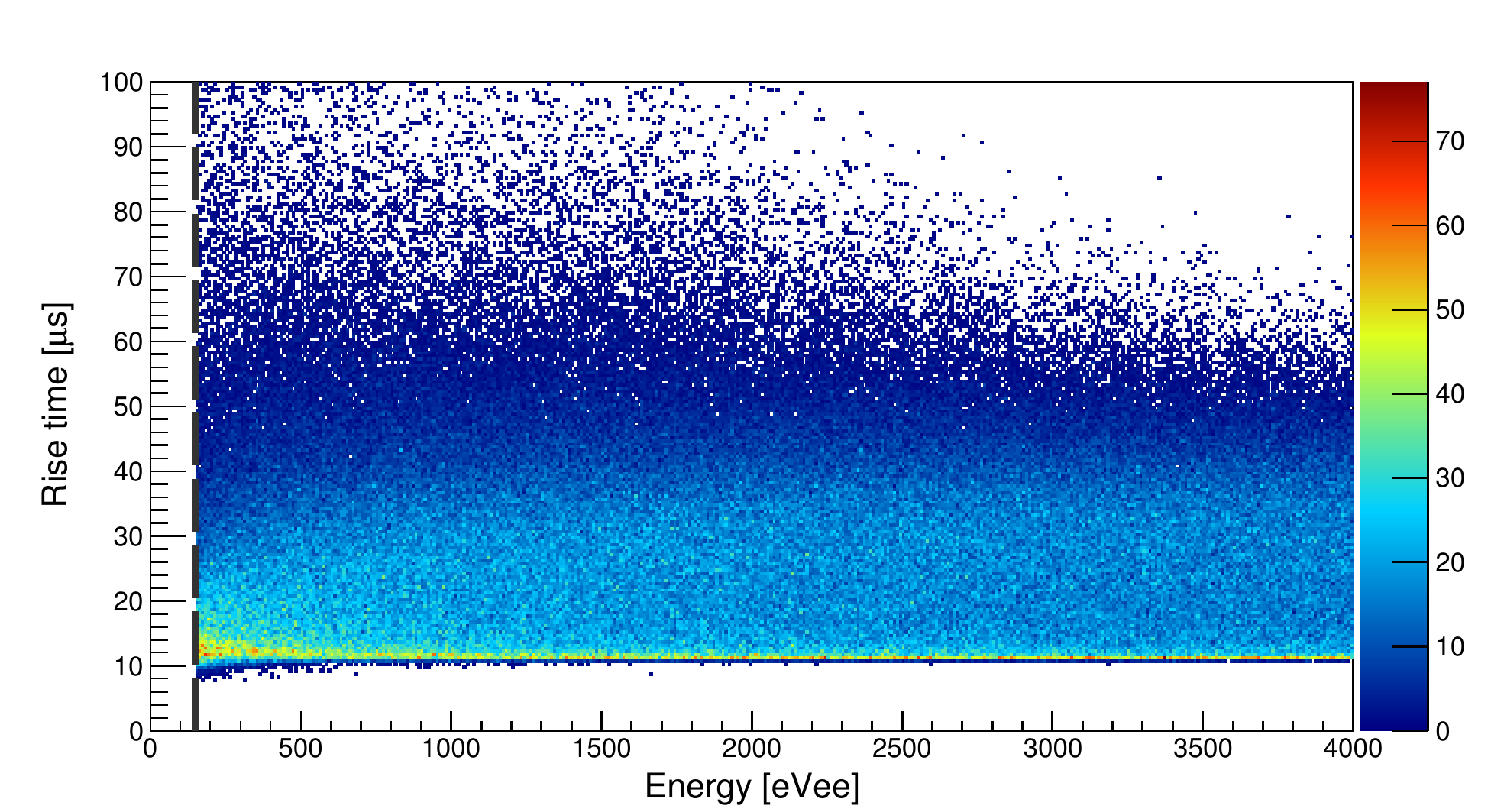}\\
\includegraphics[width=0.8\linewidth]{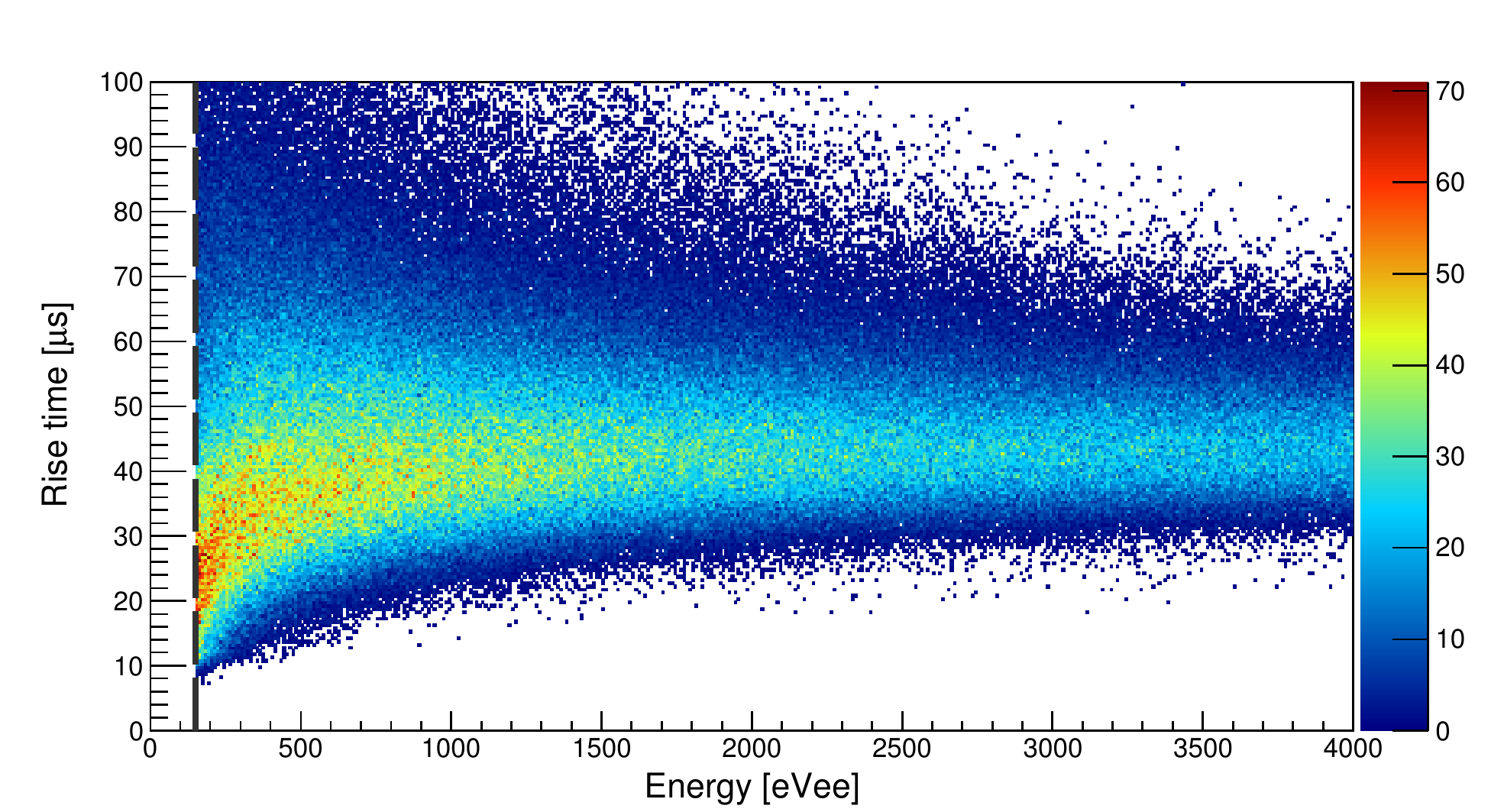}
\caption{Distribution of $10^6$ simulated bulk events (top panel) and surface events (bottom panel) in the rise time vs. energy plane, obtained from a Monte Carlo simulation described in section \ref{sec:simulations}. The statistical discrimination against surfaces events from bulk events is enabled down to energies as low as the analysis threshold of $150\;\mathrm{eV_{ee}}$ indicated in dashed black line.}
\label{fig:simulations}
\end{figure}
In Fig.~\ref{fig:simulations} we show the distribution of $10^6$ simulated bulk events (top panel) and surface events (bottom panel) in the rise time vs. energy plane, obtained from a Monte Carlo simulation described in section~\ref{sec:simulations}. For bulk events, the rise time distribution is rather uniform with the exception of the visible over-density around $10\;\mathrm{\mu s}$. The latter is a consequence of the single deconvolution procedure which prevents the reconstruction of events at a lower rise time. Surface events whose diffusion time is $\sim21\;\mathrm{\mu s}$ are peaked around $\mathrm{RT}=42\;\mathrm{\mu s}$ at high energy whereas the simple linear dependence $\mathrm{RT}=1.96\times \sigma$ no longer holds at low energy due to the low statistics of PEs per pulse.  Still, the rise time provides useful statistical discrimination against surface events down to our analysis threshold of $150\;\mathrm{eVee}$ indicated by the dashed black line.
\section{Simulation and Calibration}
\label{sec:simulations}
In this section we describe the simulation used to determine our sensitivity to WIMPs. We make use of neutron calibration data from an Am-Be source and an \textit{in situ} low-energy $X$-ray $^{37}$Ar gaseous source to validate our signal modelling. 
\subsection{Calibration}
\label{subsec:Calibration}
The $^{37}$Ar gaseous source was produced by irradiating pure $^{40}$Ca powder with fast neutrons from an Am-Be source. $^{37}$Ar gas was added to the mixture of Ne + $\mathrm{CH}_4$ at the end of the run, providing a large sample of mono-energetic events at 2.82 keV and 270 eV from X-rays induced by electron capture in the K- and L-shells, respectively~\cite{Ar37recent,Ar37dougan,Ar37santos,Ar37mostcited}. The 2.82 keV peak in the ${}^{37}\mathrm{Ar}$ calibration data exhibits substantial energy losses linearly increasing with the rise time. As the pulse treatment efficiently corrects for ballistic deficit, these are attributed to electron attachment. By fitting the 2.82 keV peak in the energy vs. rise time plane, we determined the correction function to apply to the data to correct the reconstructed energy for electron attachment. The neutron calibration provides homogeneously distributed events (confirmed from Geant4 simulations) down to the critical low energy range of $[150,250]\;\mathrm{eV_{ee}}$ where our sensitivity to sub-$\mathrm{GeV/c^2}$ WIMPs comes from. It allows us to verify the response of the simulation in rise time by comparing the calibration to simulated events distributed homogeneously in the detector.
\subsection{Simulation}
\label{subsec:simu}
For electronic recoils, the mean energy to create one electron-ion pair in neon gas is $W_\gamma=36\;\mathrm{eV}$ down to relatively low recoil energy $E_R$, but it increases asymptotically when getting close to the ionization potential $\mathrm{U}=21.6\;\mathrm{eV}$. We considered the theoretically motivated parametrization $W_\gamma(E_R)= (E_R/(E_R-21.6))\times 36 \;\mathrm{eV}$~\cite{Wvalue1} confirmed from measurements \cite{Wvalue2,WvalueCombecher}. For nuclear recoils, we derived the quenching parametrization $Q(E_R)=0.216 \times (E_R)^{0.163}$ from a Monte Carlo simulation code called SRIM (Stopping and Range of Ions in Matter)~\cite{srim} and conservatively considered $W_n(E_R)=W_\gamma(E_R)/Q(E_R)$. A field map of the full detector geometry was computed using finite-element software to account for field inhomogeneity effects due to the presence of the rod (see Fig~\ref{fig:detector}). 

The PEs were individually drifted according to electric field-dependent diffusion coefficients and drift velocities determined with a CERN simulation package called Magboltz~\cite{magboltz}. The electron attachment rate was parametrized as a function of the field based on the rise time dependence of the energy losses observed in the ${}^{37}\mathrm{Ar}$ calibration run. PEs not reaching the sensor (e.g. ending on the wire insulator or on the surface of the sphere) were considered lost ($\sim$13~\% of the volume is affected by partial loss of PEs). The avalanche process was simulated with Garfield++~\cite{Garfieldpp} which accounts for Penning transfers~\cite{PenningNe,PenningAr,PenningGarfield}: processes by which excitation energy of atoms ($\mathrm{Ne}^{*}$) is used to ionize the admixture molecules ($\mathrm{CH_4}$), significantly increasing the amplification gain. Because of the high computing time required to simulate the avalanche, it was only performed once for $3\times10^4$ PEs, and the distribution of the number of secondary ionizations, $n$, fitted with the Polya distribution~\cite{polya-Alkhazov,spclow}:
\begin{equation}
P (\frac{n}{\langle n \rangle})	=\frac{(1+\theta)^{(1+\theta)}}{\Gamma(1+\theta)}{\left(\frac{n}{\langle n \rangle}\right)}^{\theta}\mathrm{exp}\left[ -(1+\theta) \frac{n}{\langle n \rangle}   \right]
\label{eq:polya}
\end{equation}
where $\langle n \rangle$ is the mean gain, and $\theta$ is a parameter driving the shape of the distribution, from an exponential ($\theta=0$) to a normal distribution ($\theta\gg 1$). The value of $\theta=0.25$ that best fits the single electron response simulated with Garfield was used to draw the number of secondary electron/ion pairs from Eq~\ref{eq:polya}. 
The mean amplification gain was defined as a function of the arrival angle of the PEs, in order to account for the effect of electric-field anisotropies on the avalanche process. The angular dependence of the gain was parametrized in order to obtain good agreement between simulation and ${}^{37}\mathrm{Ar}$ data. The pulse was constructed by summing the detector response associated with each primary electron reaching the sensor. A baseline randomly chosen from empty pre-traces of events recorded during the physics-run was then added to simulate realistic noise. Simulated pulses then went through the same trigger algorithm, processing and energy correction for electron attachment as real pulses.  
\begin{figure}[t!]
\centering
\includegraphics[width=0.8\linewidth]{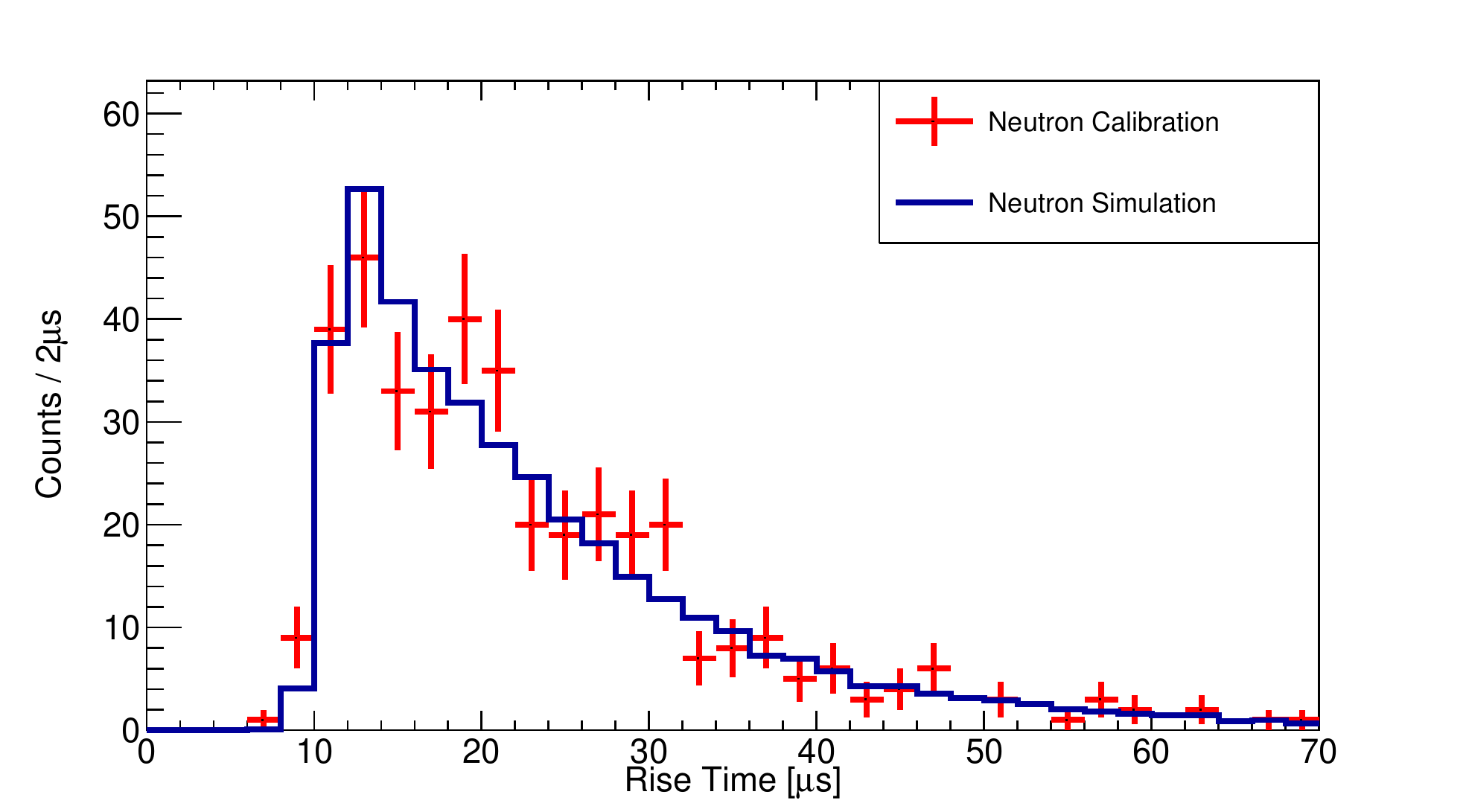}]
\caption{Agreement between the rise time distribution of events in the energy range $[150,250]\;\mathrm{eV_{ee}}$ for events from neutron calibration data (red markers) and for simulated events with energy deposits homogeneously distributed in the volume (dark blue histogram). }
\label{fig:neutronRT}
\end{figure}  
\subsection{Validation of the simulation}
We show in Fig.~\ref{fig:neutronRT} the rise time spectrum in the $[150,250]\;\mathrm{eV_{ee}}$ energy range of events recorded during the neutron calibration run with the Am-Be source (red markers). For comparison, the rise time distribution of simulated bulk events (blue histogram) is scaled to the same number of events. The good agreement, which extends to the whole energy range considered for the analysis, validates both the response of the simulation in rise time and the drift parameters derived from Magboltz. 
\begin{figure}[t!]
\centering
\includegraphics[width=0.8\linewidth]{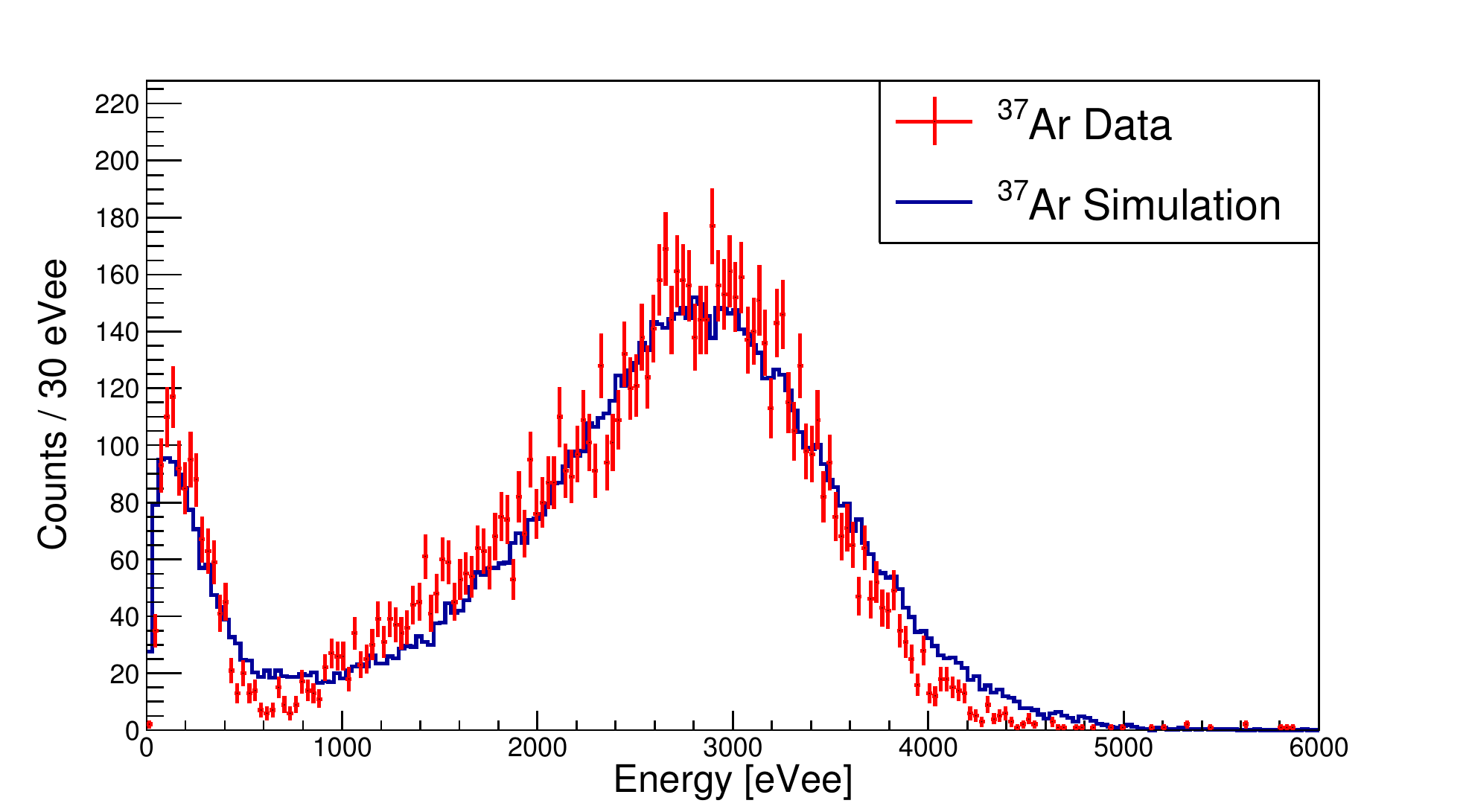}
\caption{Comparison of ${}^{37}\mathrm{Ar}$ calibration data and simulation. The measured energy spectrum (red markers) exhibits two peaks at 2.82 keV and 270 eV from the electron capture in the K- and L-shell, respectively. $10^5$ mono-energetic electronic recoils were simulated with the expected intensity ratio of 0.1 for L/K electron capture. The energy spectrum derived from the simulation (dark blue histogram) is scaled to the number of events recorded during the calibration run for comparison.  }
\label{fig:calib}
\end{figure}
In Fig.~\ref{fig:calib} we compare the energy spectra of events recorded during the ${}^{37}\mathrm{Ar}$ calibration run together with simulated 270 eV and 2.82 keV electronic recoils in the expected ratio of 0.1 for L/K electron capture. One can see that there is not only good agreement in the critical low energy range but also that the simulation is able to reproduce the non-Gaussianity of the 2.82 keV peak caused by field anisotropies in the avalanche region. The overall agreement between the simulation and calibration data allows us to confidently derive our sensitivity from simulated WIMP events.
\section{Physics-run data analysis}
\label{sec:dataanalysis}
The energy calibration was performed using the 8 keV line from the copper fluorescence in the physics-run. The energy measurement was corrected for electron attachment as described in section \ref{subsec:Calibration}. A conservative analysis threshold was set at $150\;\mathrm{eV_{ee}}$, far above the trigger threshold of $\sim36\;\mathrm{eV_{ee}}$, hence ensuring a $\sim100\;\%$ trigger efficiency even for surface events which are corrected the most for electron attachment. 
\begin{figure}[b!]
\centering
\includegraphics[width=0.8\linewidth]{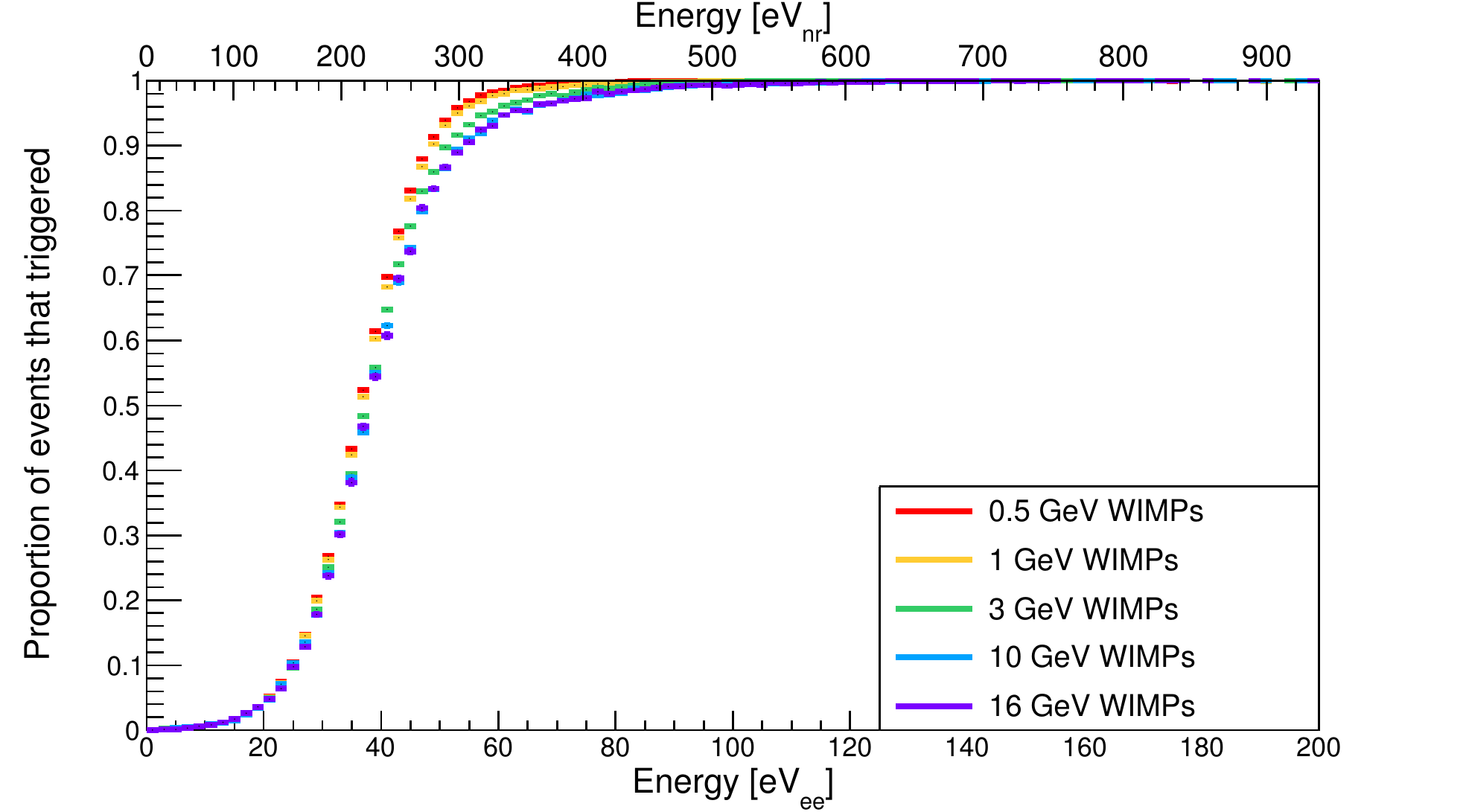}
\caption{Proportion of the events that trigger as a function of the energy of the simulated pulse for different WIMP masses as indicated by the colour code. For a given number of secondary ions, the trigger algorithm performs slightly better on single PE events, which is why the efficiency curves obtained from simulated pulses with recoil energies drawn from energy spectra of $0.5\;\mathrm{GeV/c^2}$ and $1\;\mathrm{GeV/c^2}$ WIMPs (mostly single PE events), are slightly above those derived from higher WIMP masses. For reference, a fit of the efficiency curve of $0.5\;\mathrm{GeV/c^2}$ WIMPs with $f(E)=0.5*\left[1+\mathrm{erf}\left((E-E_{th})/(\sqrt{2}*\sigma)\right)\right]$ gives  $E_{th}=36.5\;\mathrm{eV_{ee}}$ and $\sigma=9.6\;\mathrm{eV_{ee}}$}
\label{fig:trigger}
\end{figure}
We show in Fig~\ref{fig:trigger} the proportion of simulated events that trigger when pulses are added on top of a baseline (cf. section~\ref{subsec:simu}), as a function of the reconstructed energy from the pulse. This trigger efficiency was derived from simulated events of various WIMP masses to point out its dependence on the recoil energy spectrum. Notably, for a given number of secondary ions and hence for a given reconstructed energy, the trigger algorithm performs slightly better on single PE events than on multiple PE events. As a result, the efficiencies derived from simulated WIMPs of $0.5\;\mathrm{GeV/c^2}$ and $1\;\mathrm{GeV/c^2}$ masses (mainly single PE events) are slightly higher than those of higher masses. Energy scales are given both in $\mathrm{eV_{ee}}$ and in $\mathrm{eV_{nr}}$ (nuclear recoil equivalent energy) using the quenching factor discussed in section~\ref{subsec:simu}. In spite of the analysis threshold corresponding roughly to $720\;\mathrm{eV_{nr}}$, we are sensitive to sub-$\mathrm{GeV/c^2}$ WIMPs because large upper fluctuations of the avalanche gain occur with a high probability: single electrons events have for example a $\sim1.2\;\%$ chance (c.f. Eq~(\ref{eq:polya})) to produce 4 times the mean number of secondary ions throughout the avalanche process and hence to produce a signal of similar amplitude than an average 4 PE event. In addition, Poisson upper-fluctuations on the expected number of created PEs also plays a role in our sensitivity to WIMP masses in the $\mathrm{GeV/c^2}$ range. 

Quality cuts were performed on the shape of the raw pulses in order to remove non-physical events such as micro-discharges that exhibit raw pulses whose shape corresponds to the preamplifier response function only. The applied cuts are loose enough not to lead to any other signal efficiency loss than the dead time they induce. Spurious ``after pulses'' were sometimes observed after a physical event. These pulses have high enough amplitudes to trigger the readout and are difficult to reject on a shape-basis. To remove these non-physical events we reject events occurring within a 4 seconds time-window after triggering. The dead time associated with these quality cuts leads to a 20.1~\% exposure loss.
\begin{figure}[b!]
\centering
\includegraphics[width=0.8\linewidth]{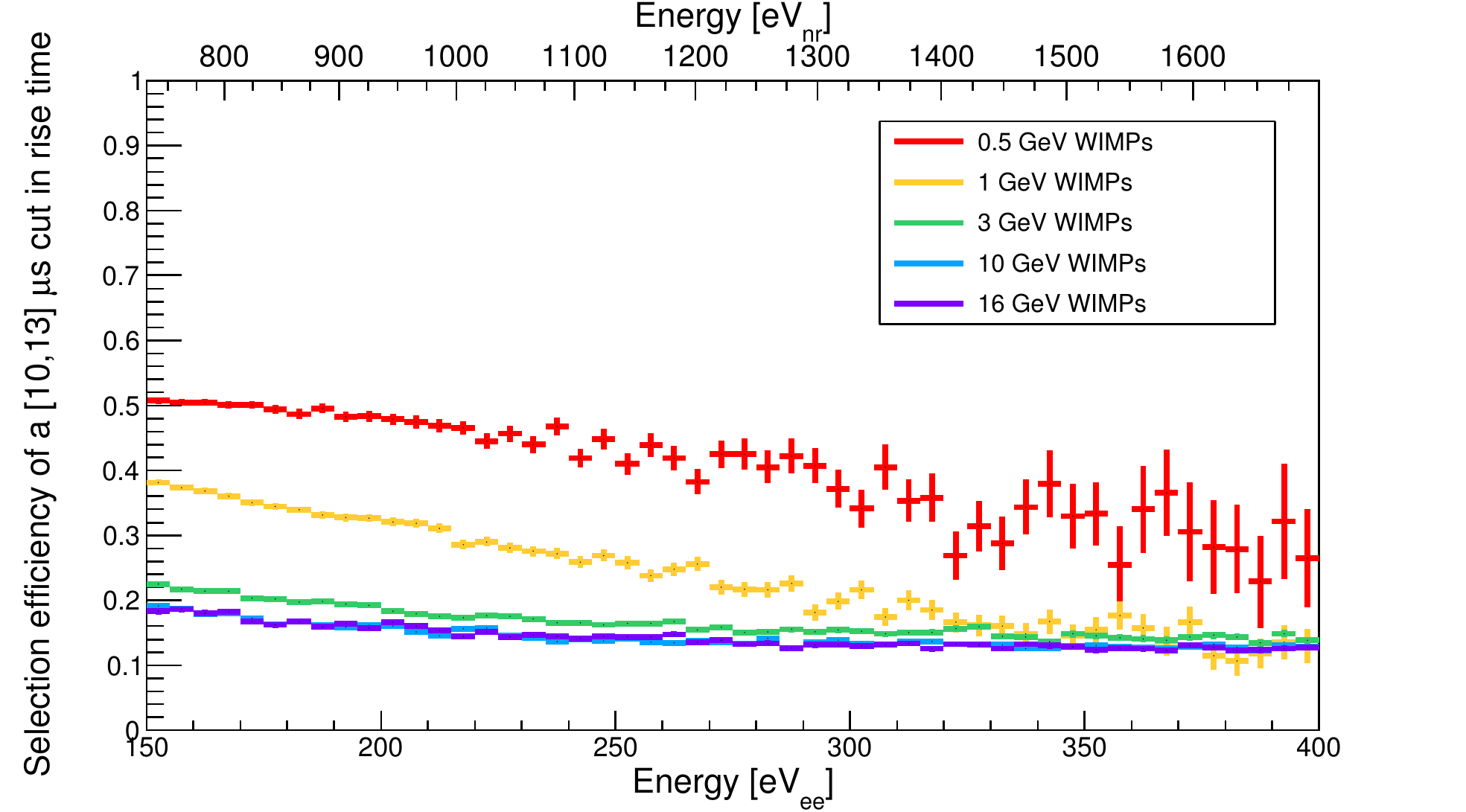}
\caption{Selection efficiency of a cut that keeps events in the rise time range of 10 to $13\;\mathrm{\mu s}$, as a function of the reconstructed energy for simulated WIMPs of different masses as indicated by the colour code. Binomial error bars get larger for low-mass WIMPS as the energy increases because of the decreasing number of simulated events to derive the efficiency. The rise-time cut efficiency decreases as the WIMP mass increases until an asymptotic regime is reached, after which the slope of the decaying spectrum no longer affects the rise time distribution.}
\label{fig:rtcutefficiency}
\end{figure}

Counterintuitively, the signal efficiency of a cut in rise time depends on the WIMP mass considered in the low energy range. For a cut that keeps events in the rise time range of 10 to $13\;\mathrm{\mu s}$, which is typically the cut that optimizes our sensitivity to sub-$\mathrm{GeV/c^2}$ WIMPs, we show on Fig~\ref{fig:rtcutefficiency} that the proportion of simulated WIMP events that pass the cut decreases as the WIMP mass increases. To explain this behaviour, we consider the following illustrative example: when an event is reconstructed at $150\;\mathrm{eVee}$, it is most likely from a $16\;\mathrm{GeV/c^2}$ WIMP (resp. $0.5\;\mathrm{GeV/c^2}$), which exhibits a slow- (resp. fast-) decaying energy spectrum, that the signal originates from 5 PE (resp. 1 PE) with a mean gain from the avalanche of $\sim \langle n\rangle$ (resp. $\sim5\times \langle n\rangle$). Single PE events will result in rise times lower than the sampling period, still artificially reconstructed around 10 $\mu s$ due to the pulse treatment, whereas for multiple PE events, the rise time will increase with the radial location of the event. As a consequence, it is not possible to use either the Am-Be, nor the ${}^{37}\mathrm{Ar}$ calibration data to determine the signal efficiency of cuts in rise time for low-mass WIMPs who have a sharp recoil energy spectrum. However, for flat and slow-decaying spectra, the rise time distribution of events is almost independent of the recoil energy spectrum, as one can see in Fig~\ref{fig:rtcutefficiency} from the overlap of the rise-time selection efficiency curves for $10\;\mathrm{GeV/c^2}$ (blue line) and $16\;\mathrm{GeV/c^2}$ (purple line) WIMPs. Thus, the Am-Be calibration whose energy spectrum decays slowly can be used to assess the validity of the simulation, without relying on any assumption of its underlying recoil energy spectrum. The good agreement between simulations and Am-Be calibration data, discussed in section~\ref{fig:simulations} and shown in Fig~\ref{fig:neutronRT} allows us to confidently rely on the simulation to determine our rise-time cut efficiencies.

For the analysis, a wide preliminary Region Of Interest (ROI) was defined in rise time $[10,32]\;\mathrm{\mu s}$ and energy $[150,4000]\;\mathrm{eV_{ee}}$. Side band regions were used to determine the expected backgrounds in the preliminary ROI. The event rate measured in the $[4000,6000]\;\mathrm{eV_{ee}}$ energy range was used to extrapolate the expected Compton background down to lower energy assuming a flat recoil energy spectrum. The side band region above $32\;\mathrm{\mu s}$ in rise time was used together with our simulation to extrapolate the expected number and distribution of surface events leaking in the preliminary ROI. To ensure both an optimized and conservative (discussed below) upper limit on the WIMP-nucleon cross section, further tuning of the ROI was performed using a Boosted Decision Tree~\cite{TMVA}. The latter is a multivariate analysis technique commonly used in high energy physics and more recently in dark matter searches~\cite{PANDAX,List-Limits-supercdms2014.txt,EDWIIIlowmass}. A BDT is a machine learning algorithm trained to classify events using a set of discriminating variables. It reduces the parameter space, in our case (energy, rise time), to only one variable, the BDT score, whose value ranges between -1 and 1 depending on whether the event is background-like or signal-like, respectively. 
The BDT was trained with $10^5$ simulated of both background events, with ratios determined by the side-band extrapolations, and signal events for 8 different WIMP masses (from 0.5 to 16 $\mathrm{GeV/c^2}$). For each WIMP mass, the BDT cut to be applied to the data was determined in order to optimize the expected sensitivity from the background model only. This procedure results in a WIMP-mass-dependent fine-tuned ROI in the rise time vs. energy plane.
 
\begin{figure}[t!]
\centering
\includegraphics[width=0.8\linewidth]{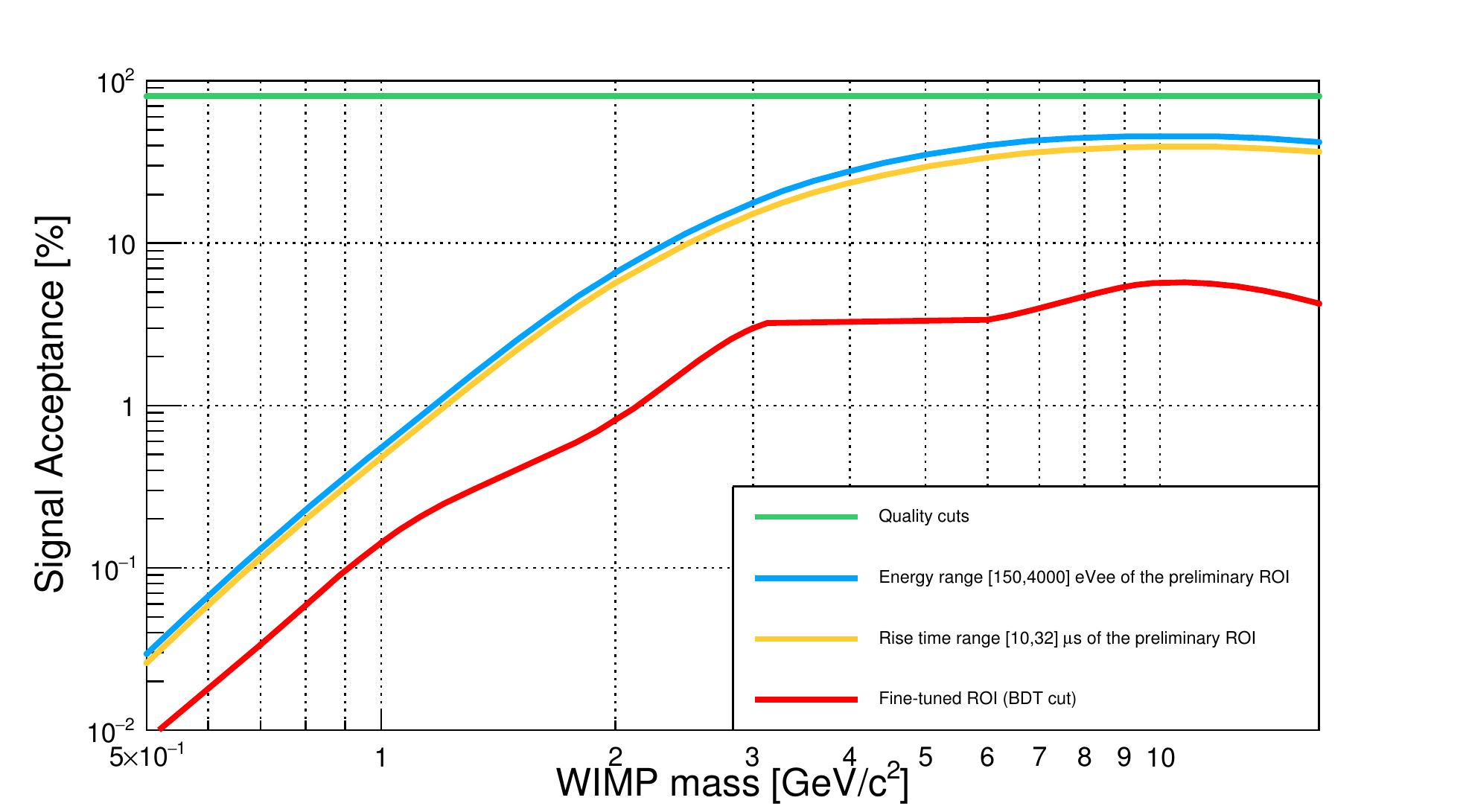}
\caption{Proportion of the simulated WIMPs that pass a successive set of cuts as a function of their mass. The green line corresponds to the quality cuts. The blue line includes the signal efficiency of a cut in energy requiring the reconstructed energy to be in the energy range $[150,4000]\;\mathrm{eV_{ee}}$ of the preliminary ROI. The orange line further includes the cut in rise time $[10,32] \;\mathrm{\mu s}$ of the preliminary ROI. Finally, the red line shows the signal survival probability of the final cut on the BDT score.}
\label{fig:signalacceptancevswimpmass}
\end{figure}
We show in Fig~\ref{fig:signalacceptancevswimpmass} the signal acceptance as a function of the WIMP mass after successive application of cuts: the blue curve indicates the proportion of simulated WIMPs with a reconstructed energy in the range $[150,4000]\;\mathrm{eV_{ee}}$ of the preliminary ROI. The signal acceptance shown in orange further includes the cut in rise time $[10,32] \;\mathrm{\mu s}$ of the preliminary ROI. Lastly, the red curve shows the signal survival probability of the final cut on the BDT score. These acceptance curves take into account the dead time induced by the quality cuts (indicated by the green line). One can see from the increasing ratio of the red curve to the blue one that the efficiency of the BDT cut decreases as the WIMP mass increases, from 29~\% for a $0.5 \;\mathrm{GeV/c^2}$ WIMP mass to 10~\% for a $16\;\mathrm{GeV/c^2}$ WIMP mass.
An interesting feature of the BDT approach is the robustness of the analysis against background mis-modeling. Indeed, if the BDT were to be trained with inaccurate background models, the fine-tuned ROI would not be optimized for signal / background discrimination, resulting in a weaker, but conservative, upper-limit set on the WIMP-nucleon cross section.
\section{Results and discussion}
\label{sec:results}
\begin{figure}[t!]
\centering
\includegraphics[width=0.9\linewidth]{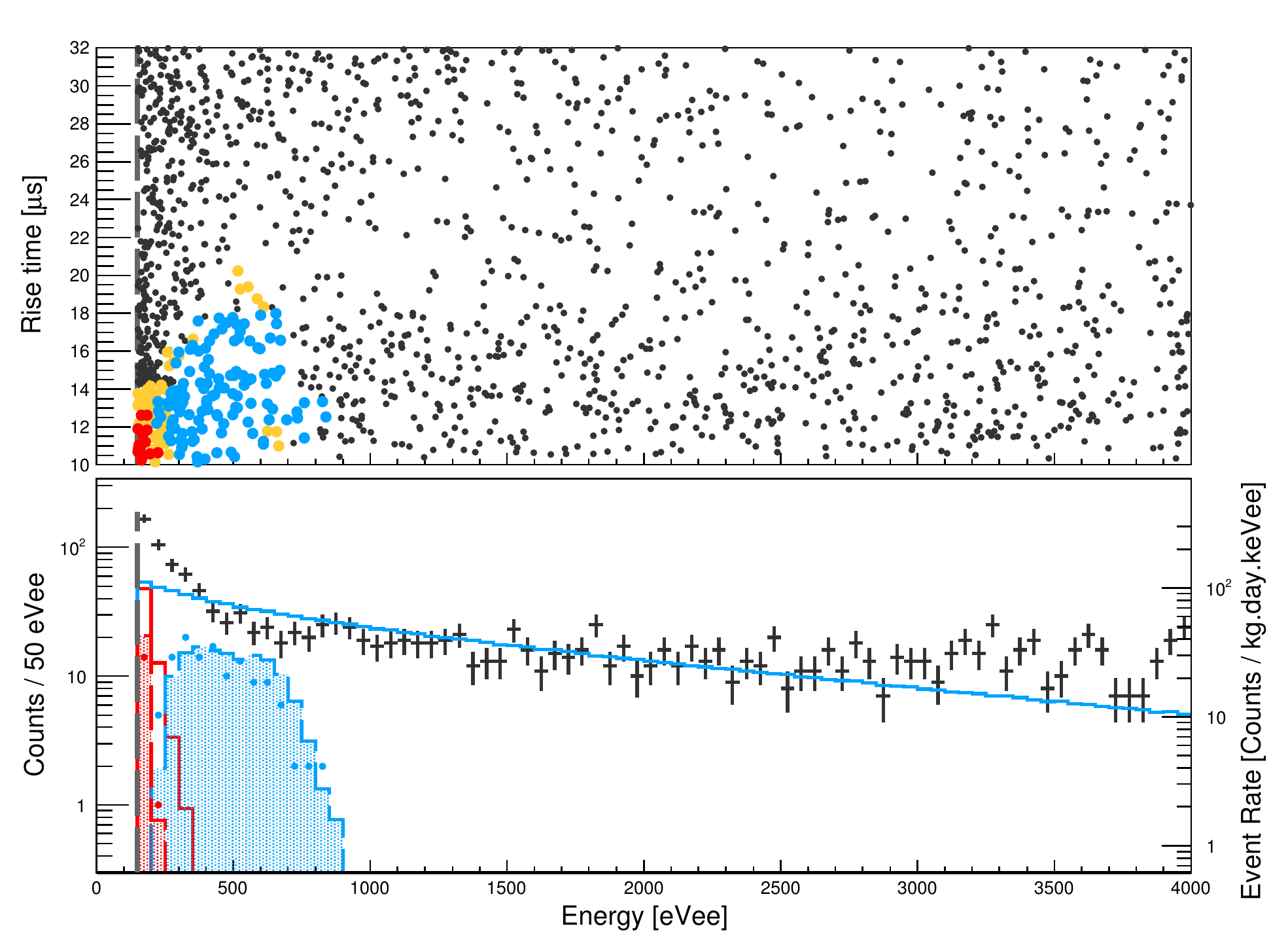}
\caption{Top panel: distribution of the 1620 events recorded during the physics run in the preliminary ROI. Events that fail (resp.~pass) the BDT cut for any of the WIMP masses are shown in black (resp. colour) dots. Events accepted as candidates for $0.5\;\mathrm{GeV/c^2}$ and $16\;\mathrm{GeV/c^2}$ WIMP masses are shown in red and blue, respectively, while for intermediate WIMP masses, candidates are shown in orange. Bottom panel: the energy spectrum of events recorded during the physics run in the preliminary ROI is indicated by the black markers. Energy spectra of $0.5\;\mathrm{GeV/c^2}$ and $16\;\mathrm{GeV/c^2}$ WIMP candidates are shown in red and blue dots. The energy spectra before and after the BDT cut of simulated $0.5\;\mathrm{GeV/c^2}$ (resp. $16\;\mathrm{GeV/c^2}$) WIMPs of cross section $\sigma_{excl}=4.4\times10^{-37}\;\mathrm{cm^2}$ (resp. $\sigma_{excl}=4.4\times10^{-39}\;\mathrm{cm^2}$) excluded at 90~\% (C.L.) are shown in solid and dashed filled red (resp. blue) histograms,  respectively.}
\label{fig:paperdual}
\end{figure}

For a total exposure of 34.1 live-days$\;\times\;0.283\;\mathrm{kg}= 9.7\;\mathrm{kg\cdot days}$ before rise-time cut efficiency correction, 1620 events were recorded in the preliminary ROI. The energy spectrum and scatter plot of these events are shown in the bottom and top panels of  Fig~\ref{fig:paperdual} as black markers and black dots, respectively.
Events that pass the BDT cut for any of the WIMP masses are shown in colour dots on top panel. To highlight the clear difference in the optimization of the ROI for sub-$\mathrm{GeV/c^2}$ and higher WIMP masses, we show in red dots (resp. blue dots) both the scatter plot and energy spectra of the 15 events (resp. 123 events) that pass the cut on the BDT score for a $0.5\;\mathrm{GeV/c^2}$ (resp.~$16\;\mathrm{GeV/c^2}$) WIMP mass. The energy spectrum of simulated $0.5\;\mathrm{GeV/c^2}$ (resp. $16\;\mathrm{GeV/c^2}$) WIMPs of cross section $\sigma_{excl}=4.4\times10^{-37}\;\mathrm{cm^2}$ (resp. $\sigma_{excl}=4.4\times10^{-39}\;\mathrm{cm^2}$) excluded at 90~\% (C.L.) is shown in red (resp. blue), both before and after the BDT cut by the solid and dashed filled histograms, respectively. One can see from the comparison of the energy spectra of simulated WIMPs with data, before and after performing the BDT cut, how the optimization of the ROI improves the sensitivity to WIMPs, even though no background subtraction is performed.

\begin{figure}[b!]
\centering
\includegraphics[width=0.9\linewidth]{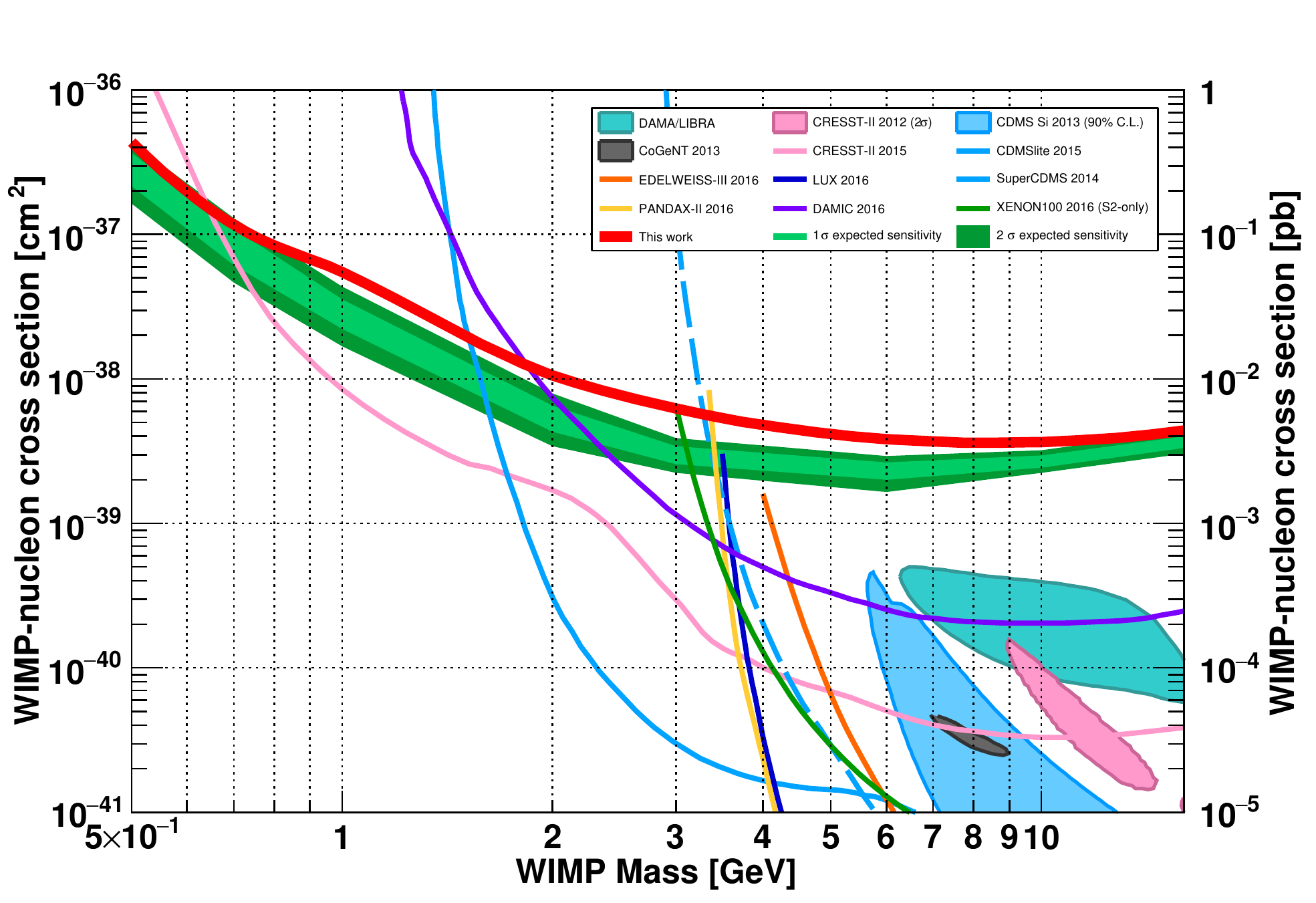}
\caption{Constraints in the Spin-Independent WIMP-nucleon cross section vs. WIMP mass plane. The result from this analysis is shown in solid red together with the expected $1\;\sigma$ (resp. $2\sigma$) sensitivity from our background-only model in light green (resp. dark green). Signal hints reported by the CDMS-II Si~\cite{List-Limits-contoursicdms90.txt}, CoGeNT~\cite{List-Limits-cogentcontour2013.txt}, DAMA/LIBRA~\cite{damalibra1of2,damalibra2of2} and CRESST-II phase 1 \cite{List-Limits-cresstcontours.txt} experiments are shown in colour contours. Results reported as an upper limit on the WIMP-nucleon cross section are shown in solid and dashed lines for the following experiments: DAMIC~\cite{DAMIC2016}, LUX~\cite{List-Limits-lux2016.txt}, XENON100~\cite{xenon2016lowmass}, CRESST-II~\cite{List-Limits-cresst2015.txt}, CDMSlite~\cite{List-Limits-cdmslite2015.txt}, SuperCDMS~\cite{List-Limits-supercdms2014.txt}, EDELWEISS~\cite{List-Limits-edelweiss2011.txt} and PANDAX-II~\cite{PANDAX}.}
\label{fig:limit}
\end{figure}

For each WIMP mass and considering as candidates all the events observed in the corresponding fine-tuned ROI, a 90~\% Confidence Level (C.L.) upper limit on the spin-independent WIMP-nucleon scattering cross section was derived using Poisson statistics. The recoil energy spectrum used to derive the sensitivity to WIMPs is based on standard assumptions of the WIMP-halo model, with a local dark matter density of $\rho_{DM}=0.3\;\mathrm{GeV/c^2/cm^3}$, a galactic escape velocity of $v_{esc}=544\;\mathrm{km/s}$, and an asymptotic circular velocity of  $v_{0}=220\;\mathrm{km/s}$. We show in Fig.~\ref{fig:limit} the resulting exclusion limit as a solid red line together with the expected $1\;\sigma$ (resp. $2\;\sigma$) sensitivity region in light green (resp. dark green) from our data-driven background model in the absence of a signal. Over all of the WIMP mass range, the exclusion limit lies either in the upper $2\;\sigma$ sensitivity region or above. This indicates that background levels have been underestimated in the fine-tuned ROI and that the latter may not be perfectly optimized for signal/background discrimination. Still, competitive constraints are set in the $\mathrm{GeV/c^2}$ mass range, and a new region of the parameter space is probed below $0.6\;\mathrm{GeV/c^2}$.

We have considered how several effects and systematic uncertainties affect the limit. For instance, a $\pm5$~\% uncertainty was attributed to the energy scale based on our determination of the position of the 8~keV fluorescence peak as well as our determination of the 0 point energy using pre-trigger traces. This results in a 17~\% increase (resp. 26~\% decrease) of our sensitivity to $0.5\;\mathrm{ GeV/c^2}$ WIMPs and has an effect of $\sim$1~\% for WIMP masses higher than $6\;\mathrm{ GeV/c^2}$. Furthermore, a population of quenched 5.3~MeV ${}^{210}\mathrm{Po}$ alphas showed a time-dependent linear shift in reconstructed energy of 3~\% over the 42.7 days of data-taking. The effect of this possible gain shift (effectively 1.5~\%) is negligible on the sensitivity when compared to the 5~\% uncertainty in the energy scale and was thus not extrapolated to the data at low energies. Nuclear quenching factor measurements in neon and helium gas are currently being undertaken by the NEWS-G collaboration down to sub-keV energies using ion beams. As this study is ongoing, we relied on SRIM simulations for our parametrization of the quenching as existing low-energy measurements have been solely performed in the keV range and with a He target~\cite{QF1,QF2}. For reference, for a $0.5\;\mathrm{ GeV/c^2}$ (resp. $3\;\mathrm{ GeV/c^2}$) WIMP mass, a decrease of 10~\% of the nuclear quenching factor translates into a 11~\% (resp. 4~\%) impact on the sensitivity. Finally, for increasing values of the Polya parameter $\theta$ (c.f. Eq.~(\ref{eq:polya})), higher fluctuations of the avalanche gain occur with a lower probability, reducing the sensitivity to the lower WIMP masses. Furthermore, higher Penning transfer rates in the gas increase the value of $\theta$ that best fits the single electron response simulated with Garfield, up to $\theta=0.3$, for which the limit would be shifted up by $7\;\%$ for a $0.5\;\mathrm{ GeV/c^2}$ WIMP mass.
\section{Conclusion and outlook}
\label{sec:conclusion}
The NEWS-G experiment sets new constraints on the spin-independent WIMP-nucleon scattering cross-section below $0.6\;\mathrm{GeV/c^2}$ and excludes at 90~\% confidence level (C.L.) a cross-section of $4.4 \times \mathrm{10^{-37}\;cm^2}$ for a 0.5 $\mathrm{GeV/c^2}$ WIMP mass. We thereby demonstrate the high potential of Spherical Proportional Counters for the search of low-mass WIMPs. Further operation of SEDINE with He gas will allow for the optimization of momentum transfers for low-mass particles in the $\mathrm{GeV/c^2}$ mass range, and increase our sensitivity to sub-$\mathrm{GeV/c^2}$ WIMPs. The next phase of the experiment will build upon the knowledge acquired from the operation of the SEDINE prototype at the LSM. It will consist of a 140 cm diameter sphere capable of holding gas up to a pressure of 10 bars, to be installed in SNOLAB by summer 2018. The sphere will be shielded by a shell of 25 cm of both archeological and low activity lead, itself inside a 40 cm thick polyethylene shield. Space in SNOLAB has been assigned, the design of the whole project is completed and technical design review is ongoing. Among many major improvements, selection of extremely low activity copper (in the range of a few $\mu$Bq/kg of U and Th impurities) and dedicated handling to avoid radon entering the detector at any time will ensure significant reduction of the backgrounds levels, both in surface and volume, relative to the above results, and allow sensitivity down to cross sections of $\mathcal{O}(10^{-41}\;\mathrm{cm^{2}})$. Use of H and He targets will allow us to reach WIMP mass sensitivity down to 0.1 GeV.
\section*{Acknowledgments}
The help of the technical staff of the Laboratoire Souterrain de Modane is gratefully acknowledged. The low activity prototype operated in LSM has been partially funded by the European Commission astroparticle program ILIAS (Contract R113-CT-2004- 506222). This work was undertaken, in part, thanks to funding from the Canada Research Chairs program, as well as from the French National Research Agency (ANR-15-CE31-0008).
\section*{References}

\bibliography{mybibliography}

\end{document}